\documentclass[12pt]{article}
\usepackage{enumerate}
\usepackage[margin=1in, top=0.8in]{geometry}
\usepackage[OT1]{fontenc}
\usepackage{amsmath,amssymb}
\usepackage{mathtools}
\usepackage{natbib}
\usepackage[usenames]{color}
\usepackage{dsfont}
\usepackage{pifont}
\usepackage{pgfplots}
\usepackage{smile}
\usepackage{multirow}
\usepackage{rotating}
\usepackage{enumerate}
\usepackage{esvect}
\usepackage{tikz}
\usetikzlibrary{patterns}
\usetikzlibrary{arrows}
\usepackage{color,xcolor}
\usepackage{colortbl}
\usepackage[colorlinks,
            linkcolor=red,
            anchorcolor=red,
            citecolor=blue
            ]{hyperref}
\usepackage{algorithm}
\usepackage{algorithmic}
\usepackage{cancel}
\usepackage{diagbox}
\usepackage{tcolorbox}
\usepackage{graphicx,subfig,wrapfig,diagbox,adjustbox}
\usepackage[frozencache=true,cachedir=minted-cache]{minted} 
\usepackage{makecell}
\usepackage{listings}
\usepackage{tabularx}
\definecolor{DarkGreen}{RGB}{30,120,22}

\usepackage{fancyvrb}
\RecustomVerbatimCommand{\VerbatimInput}{VerbatimInput}
{fontsize=\footnotesize,
 breaklines=true,
 breakanywhere=true, 
 breaksymbol=,
 frame=single,  
 framesep=0.5em,
 labelposition=topline,
}

\definecolor{ultramarine}{rgb}{0.,0.1,0.9}

\usepackage{paralist}

\def\supp{\mathop{\text{supp}}}

\long\def\comment#1{}

\newcommand{\bel}{\begin{eqnarray}\label}
\newcommand{\eel}{\end{eqnarray}}
\newcommand{\bes}{\begin{eqnarray*}}
\newcommand{\ees}{\end{eqnarray*}}

\let\emptyset\varnothing

\let\tilde\widetilde
\def\mid{\,|\,}

\def\supp{\mathop{\text{supp}\kern.2ex}}
\def\argmin{\mathop{\text{\rm arg\,min}}}
\def\argmax{\mathop{\text{\rm arg\,max}}}

\def\supp{\mathop{\text{supp}}}

\def\##1\#{\begin{align}#1\end{align}}
\def\$#1\${\begin{align*}#1\end{align*}}

\usepackage{undertilde}
\theoremstyle{plain}

\title{DSTC: Direct Preference Learning \\ with Only Self-Generated Tests and Code\\ to Improve Code LMs}

\author{Zhihan Liu\thanks{Northwestern University. \texttt{\{zhihanliu2027,shenaozhang2028\}@u.northwestern.edu,zhaoranwang@gmail.com}} \qquad Shenao Zhang\footnotemark[1]\quad
    Yongfei Liu\footnotemark[2]\thanks{ByteDance Inc.  \texttt{\{liuyongfei.0314,boyi.liu01,yingxiang.yang\}@bytedance.com}} \\
    Boyi Liu\footnotemark[2] \qquad Yingxiang Yang\footnotemark[2]  \qquad
    Zhaoran Wang\footnotemark[1]}

\date{}

\begin{document}

\maketitle
\begin{abstract}
Direct preference learning offers a promising and computation-efficient beyond supervised fine-tuning (SFT) for improving code generation in coding large language models (LMs). However, the scarcity of reliable preference data is a bottleneck for the performance of direct preference learning to improve the coding accuracy of code LMs. 
In this paper, we introduce \underline{\textbf{D}}irect Preference Learning with Only \underline{\textbf{S}}elf-Generated \underline{\textbf{T}}ests and \underline{\textbf{C}}ode (DSTC), a framework that leverages only self-generated code snippets and tests to construct reliable preference pairs such that direct preference learning can improve LM coding accuracy without external annotations. DSTC combines a minimax selection process and test-code concatenation to improve preference pair quality, reducing the influence of incorrect self-generated tests and enhancing model performance without the need for costly reward models. When applied with direct preference learning methods such as Direct Preference Optimization (DPO) and Kahneman-Tversky Optimization (KTO), DSTC yields stable improvements in coding accuracy (pass@1 score) across diverse coding benchmarks, including HumanEval, MBPP, and BigCodeBench, demonstrating both its effectiveness and scalability for models of various sizes. This approach autonomously enhances code generation accuracy across LLMs of varying sizes, reducing reliance on expensive annotated coding datasets.
\end{abstract}

\tableofcontents

\newpage

\newtheorem{proposition}{Proposition}
\section{Introduction}
The recent advancement of large language models (LLMs) has significantly transformed numerous fields, including code generation. LLMs pre-trained on vast code corpora demonstrate substantial potential in generating code, and many studies have further enhanced these code models through supervised fine-tuning (SFT) to improve performance on coding tasks. Although SFT has shown promising results in improving code LMs, SFT focuses on replicating patterns within the dataset and often lacks the generalization needed to perform well on unseen instructions. This limitation has driven researchers to apply reinforcement learning (RL) to improve code LMs beyond the SFT stage by allowing them to adapt to more complex or variable inputs.

However, one of the main challenges in applying RL to code LMs is the scarcity of high-quality preference datasets tailored specifically to code generation tasks. Unlike standard language tasks, where user or human annotator feedback is widely available, a preference dataset for the code generation has unique requirements, including correctness and functional alignment, which complicates data collection and is much more expensive. Recent works, such as \textit{Starcoder2-15b-instruct} \citep{wei2024selfcodealign}, leverage execution feedback from the compiler to create SFT training datasets from self-generated tests and code snippets, aiming to simulate preference signals without human annotations. However, this reliance on self-generated tests introduces risks: inaccuracies in these synthetic tests may result in unreliable annotations, which can negatively impact the effectiveness of the RL training process. For instance, in our experiments using the self-generated \textit{oss-instruct-sd2-50k} dataset created by \textit{Starcoder2-15b-instruct}, only 48\% of responses are correct when evaluated with an external model, \textit{Deepseek-coder-v2-instruct} (with 236 billion parameters), highlighting a significant failure rate that suggests the need for a more reliable approach to leverage these synthetic tests effectively.

In addition, direct preference learning methods, such as Direct Policy Optimization (DPO, \cite{rafailov2024direct}) and Kahneman-Tversky Optimization (KTO, \cite{ethayarajh2024kto}), offer advantages in terms of computational efficiency and simplicity compared to more complex RL methods like PPO \citep{schulman2017proximal} as they bypass the training of a reward model and a critic model. Therefore, we select the direct preference learning method as the backbone of the RL training framework and analyze the following problem in this paper: 

\begin{center}
   \textit{With only self-generated code and tests, how can we generate reliable preference pairs to improve code generation accuracy for code LMs through direct preference learning?}
\end{center}

\begin{figure}
    \centering
\includegraphics[width=1\linewidth]{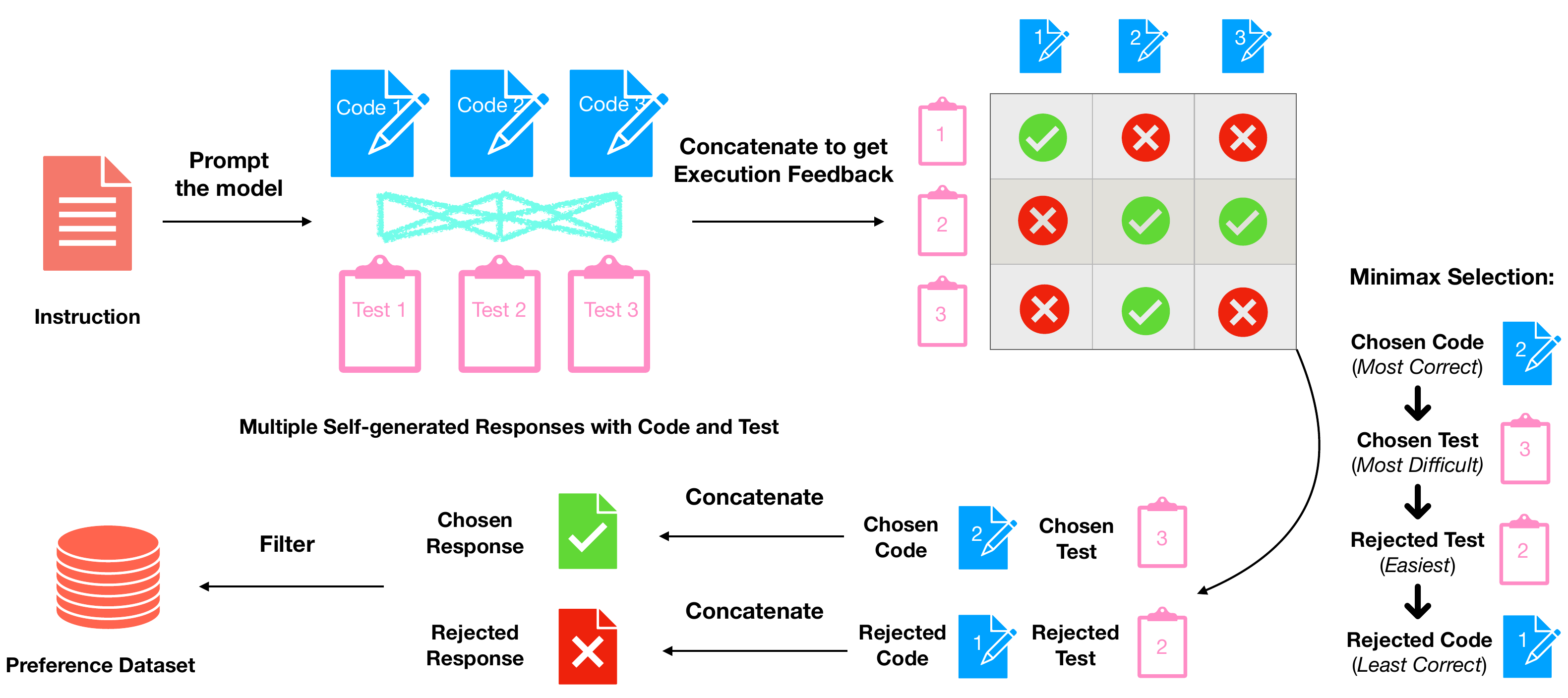}
    \caption{Illustration of our proposed mechanism DSTC to reduce the misleading effect of the low-quality test and generate a more reliable preference dataset.}
    \label{fig:illus}
\end{figure}

To address this question, we introduce a novel approach called \underline{\textbf{D}}irect Preference Learning with Only \underline{\textbf{S}}elf-Generated \underline{\textbf{T}}ests and \underline{\textbf{C}}ode (DSTC), which leverages only self-generated code snippets and tests to construct effective preference pairs. DSTC, as depicted in Figure \ref{fig:illus}, comprises two main components: (1) a minimax selection mechanism and (2) a concatenation of code snippets and tests. This framework aims to achieve two goals: (a) to enhance the quality of the chosen code-test pairs while enlarging the quality gap between the chosen and rejected code-test pair, and (b) to create more reliable preference pairs by training on concatenated code-test pairs that contain reliable binary execution feedback. By structuring the learning process in this way, DSTC reduces the noise from incorrect self-generated tests and establishes a more robust framework for preference learning in code generation tasks.
By incorporating both DPO and KTO, we implement DSTC on the model \textit{Starcoder2-15b-instruct} with 15 billion parameters and evaluate the trained model on various benchmarks involving  HumanEval Base \citep{chen2021evaluating}, HumanEval Plus \citep{liu2024your}, Mostly Basic Python Problems (MBPP) Base \citep{austin2021program}, MBPP Plus \citep{liu2024your}, and  BigCodeBench (BCB) \citep{zhuo2024bigcodebench}. 
Our results indicate that DSTC effectively increases the coding accuracy across multiple benchmarks, paving the way for designing an autonomous system to improve the ability of LM code generation. We also perform ablation studies to show the importance of each component in DSTC and demonstrate that DSTC can also improve the performance of  the model 
 \textit{Deepseekcoder-33b} with 33 billion parameters. 
\subsection{Contributions}
We summarize our contributions in two-folds.
\begin{itemize}
    \item We propose a mechanism DSTC to generate the preference pair from the self-generated code and tests to perform direct preference learning algorithms like DPO and KTO. DSTC consists of a minimax selection procedure and concatenation of the code and test. As a result, DSTC  constructs preference pairs with a reliable correct order and enlarges the quality gap between the chosen code and rejected code, which reduces the misleading effect of incorrect self-generated tests and aids the training of direct preference learning methods such as DPO and KTO.
\item Experimental results demonstrate that combining our proposed DSTC mechanism with DPO and KTO improves coding accuracy (pass@1 score) in code LMs across multiple benchmarks and model sizes (15 billion and 33 billion parameters). Ablation studies further validate the importance of each component within DSTC.
\end{itemize}

\subsection{Related Works}

\paragraph{Reinforcement Learning from Human Feedback}
Our work builds on a rich body of research on Reinforcement Learning from Human Feedback (RLHF), a framework that has become essential for training state-of-the-art large language models (LMs) such as ChatGPT \citep{achiam2023gpt}, Gemini \citep{team2023gemini}, and Claude \citep{claude}. In the RLHF pipeline, LMs are fine-tuned to optimize a learned reward model using the Proximal Policy Optimization (PPO) algorithm \citep{schulman2017proximal}. However, PPO is often criticized for its instability, sample inefficiency, and sensitivity to hyperparameter tuning \citep{engstrom2020implementation}, making it computationally expensive and less accessible for the open-source community.

To address these limitations, recent research in RLHF has explored alternatives to PPO-based methods, with direct preference learning algorithms emerging as a promising direction \citep{zhao2023slic, rafailov2024direct}. These approaches bypass the reward model training phase, aligning LMs directly with human preferences. Since the introduction of the original Direct Preference Optimization (DPO) algorithm \citep{rafailov2024direct}, numerous variants of the direct preference learning framework have been proposed, each tackling specific challenges from different angles \citep{liu2023statistical, azar2023general, xiong2023gibbs, tang2024generalized, ji2024towards, ye2024theoretical, pal2024smaug, hong2024orpo, rosset2024direct, liang2024robust, zhang2024negative, zhang2024self, zhang2024reward, xu2024just, tajwar2024preference, wu2024self, liu2024provably}.

One notable example is Kahneman-Tversky Optimization (KTO), proposed by \citet{ethayarajh2024kto}, which leverages the Kahneman-Tversky model of human utility to minimize a human-aware loss, in contrast to the standard Bradley-Terry model. While these methods prioritize improving LM general alignment abilities with human preferences, our work focuses on a different and specific goal: enhancing pass@1 scores in code generation.

\paragraph{Large Language Models for Code Generation.} Besides the great achievements in Natural Language Processing, LMs exhibit great potential in several complex reasoning tasks like math \citep{ahn2024large} and coding \citep{jiang2024survey}. In particular, pre-training on code datasets has resulted in impressive performance from models such as StarCoder \citep{lozhkov2024starcoder}, DeepSeekCoder \citep{zhu2024deepseek}. Other models, like CodeQwen \citep{hui2024qwen2} and CodeLlama \citep{roziere2023code}, have further enhanced their capabilities through ongoing pre-training with additional code data.  
To further improve the coding performance, there exists a rich line of literature that study how to fine-tune the current pre-trained LMs for coding. 
Among them, \cite{wei2024magicoder} propose a mechanism to enlighten LMs with open-source code snippets to generate diverse instruction and \cite{luo2023wizardcoder} adapt the Evol-Instruct method in the model fine-tuning. 

As another line of research to improve coding ability via RL, \cite{le2022coderl} first introduce the RL framework in the training of LMs for code generation. This RL framework is refined by PPOCoder \citep{shojaee2023execution} that incorporates the PPO algorithm in the training and \cite{liu2023rltf} refine PPOCoder with a fine-grained feedback and a Critic Sampling mechanism. To tackle the exploration issue, \cite{dou2024stepcoder} propose StepCoder by decomposing the long sequences code generation task. \cite{shen2024policy} introduce several variants of PPO by considering the policy filtration to improve the coding ability significantly. 

Our work aligns with emerging research that applies Direct Preference Optimization (DPO) and Kahneman-Tversky Optimization (KTO) to train LMs using synthetic data. For instance, \citet{gee2024code} design a method to generate preference pairs from self-generated code based on correctness and efficacy, while \citet{zhang2024plum} explore how KTO can improve coding ability with tests generated by ChatGPT.

In contrast to these methods, our approach does not rely on ground-truth tests or correct solutions, which are expensive and difficult to obtain. Instead, our method leverages the current model without intervention from stronger models like ChatGPT. Furthermore, our method improves the pass@1 score after the supervised fine-tuning (SFT) phase by integrating direct preference learning methods such as DPO and KTO, which are more computationally efficient and stable than PPO-based approaches.


\section{Preliminary}

\subsection{Reinforcement Learning from Human Feedback and Direct Policy Optimization}
Reinforcement Learning from Human Feedback (RLHF) utilizes human preferences to guide the training of language models (LMs). A common approach involves pairwise preference modeling, where feedback is provided by comparing two model-generated responses to the same prompt. Consider a language model represented as a policy $\pi_\theta(\cdot|\cdot)$, parameterized by $\theta$. Given a prompt $x$ from the state space $\mathcal{X}$, the model generates a response $a$ from the distribution $\pi_\theta(\cdot|x)$. For a specific prompt $x$, the model generates two candidate responses $a^1$ and $a^2$ from the action space $\mathcal{A}$. Human evaluators, or a reward model approximating human preferences, provide feedback as a binary label $y \in \{0, 1\}$, indicating whether $a^1 \succ a^2$ ($y = 1$) or $a^2 \succ a^1$ ($y = 0$).
This preference is modeled probabilistically using the Bradley-Terry (BT) model \citep{bradley1952rank} as follows:
\begin{equation} \mathbb{P}_r(b = 1 | x, a^1, a^2) = \sigma(r(x, a^1) - r(x, a^2)), \label{eq:bt} \end{equation}
where $r(x, a)$ is the human-provided score reflecting the quality of response $a$ for prompt $x$, and $\sigma(z) = 1/(1 + \exp(-z))$ is the sigmoid function.

The goal of RLHF is to learn a policy $\pi_\theta$ that maximizes the KL-regularized expected reward:
\begin{align} \max_{\theta} \mathbb{E}_{x \sim \cD, a \sim \pi_\theta(\cdot \mid x)} \bigl[ r(x, a) - \beta \cdot \text{KL}(\pi_\theta(\cdot \mid x) | \pi_\text{ref}(\cdot \mid x)) \bigr], \label{eq:rlhf} \end{align}
where $\pi_\text{ref}$ is the reference policy (e.g., the initial SFT model), $\beta$ is a hyperparameter, and $\cD$ represents the dataset. Standard RLHF methods to solve \eqref{eq:rlhf}, such as Proximal Policy Optimization (PPO, \citep{schulman2017proximal}), employ a two-stage process: first, a reward model is trained on preference data, and then the policy model is fine-tuned using the reward model. Recently, direct preference learning algorithms have emerged to address the RLHF problem \eqref{eq:rlhf} without training a reward model.  As the most popular one, Direct Preference Optimization (DPO) \citep{rafailov2024direct} leverages \eqref{eq:bt} and \eqref{eq:rlhf} to reparameterize the policy through the reward model, enabling direct training of the policy model on preference data.  Given the preference dataset \(\mathcal{D}_\text{DPO} = \{(x_i, a^+_i, a^-_i)\}_{i=1}^N\) of $N$ preference pairs, DPO loss has the following closed form

\begin{equation}
\mathcal{L}_{\mathrm{DPO}}\left(\theta\right)=-\mathbb{E}_{(x, a^+,a^-)\sim \mathcal{D}_\text{DPO}}\left[\log\sigma(r_\theta(x, a^+)-r_\theta(x, a^-))\right],
\label{reg_loss}
\end{equation}
where 
\begin{align}
    r_\theta(x, a) & =\log \frac{\pi_\theta(a \mid x)}{\pi_{\text {ref }}(a \mid x)}. 
\end{align}
\subsection{Kahneman-Tversky Optimization}
Instead of using the BT preference model \eqref{eq:bt} as DPO, \cite{ethayarajh2024kto} proposed another direct preference learning algorithm Kahneman-Tversky optimization (KTO) from the view of Kahneman \& Tversky’s prospect theory. In KTO, the preference dataset $\mathcal{D}_{\text{KTO}} = \{(x_i,a_i,b_i)\}_{i=1}^N$ contains $N$ pairs of a prompt $x_i$ and a response $a_i$, accompanied by a binary variable  $b_i$ that indicates whether $a_i$ is desired. Given the preference dataset $\mathcal{D}_{\text{KTO}}$ and the reference policy $\pi_\text{ref}$, KTO loss is defined as 
\begin{align}\mathcal{L}_{\mathrm{KTO}}\left(\theta\right)=\mathbb{E}_{(x, a, b ) \sim \mathcal{D}_{\text{KTO}}}\left[\lambda_y-v(x, a)\right],
\end{align}
where
\begin{equation}
\begin{aligned}
v(x, a) & =\left\{\begin{array}{l}
\lambda_D \sigma\left(\beta\left(r_\theta(x, a)-z_0\right)\right) \text {, if } b = 1 \\
\lambda_U \sigma\left(\beta\left(z_0-r_\theta(x, a)\right)\right) \text {, otherwise } \end{array}\right.\\ 
r_\theta(x, a) & =\log \frac{\pi_\theta(a \mid x)}{\pi_{\text {ref }}(a \mid x)} \\
z_0 & =\mathbb{E}_{(x^\prime, a^\prime, b^\prime ) \sim \mathcal{D}_{\text{KTO}}}\left[\operatorname{KL}\left(\pi_\theta\left(a^{\prime} \mid x\right) \| \pi_{\text {ref }}\left(a^{\prime} \mid x\right)\right) \right]. 
\end{aligned}
\end{equation}
Compared with DPO, we remark that KTO allows an imbalance between the number of chosen and rejected responses.
\subsection{Large Language Models for Code Generation}
We formulate the code generation problem in the context of RLHF, where state $x\in\mathcal{X}$ refers to some natural language instruction and action $a\in\mathcal{A}$ refers to the response provided by the code LM $\pi_\theta(\cdot\mid x)$ that contains a code snippet $y\in\mathcal{Y}$. To check the correctness of a code snippet, one common approach is to feed the concatenation of the code snippet $y$ and a test $z$ to a compiler and output the binary execution feedback. Here, test $z\in\mathcal{Z}$ can be either hand-crafted or generated by LMs, which can also be incomplete or even wrong. Since we can execute the concatenation of code and test in a compiler and receive the binary feedback (pass or fail), we define a binary reward function $r(z,y):\mathcal{Z}\times\mathcal{Y}\mapsto\{0,1\}$ for any code snippet $y$ and test $z$. Assume that there exists a ground-truth test $z^\star_x$ for the instruction $x$ such that  $z^\star_x$ can perfectly check the correctness of any code snippet for the instruction $x$. Similar to the RLHF objective in \eqref{eq:rlhf}, the objective of code generation task can be formulated as
\begin{align} \max_{\theta} \mathbb{E}_{x \sim \cD, a \sim \pi_\theta(\cdot \mid x)} \bigl[ r(z^\star_x, \texttt{Ex}(a)) - \beta \cdot \text{KL}(\pi_\theta(\cdot \mid x) | \pi_\text{ref}(\cdot \mid x)) \bigr], \label{eq:code_rlhf} \end{align}
where $\texttt{Ex}$ function extracts the code snippet $y$ from the response $a$. The similarity between \eqref{eq:rlhf} and \eqref{eq:code_rlhf} motivates us to leverage the direct preference learning method like DPO and KTO to solve the code generation task while bypassing the training of a reward function.  

\section{Method}
\label{sec:method}
\paragraph{Overview.}
In this section, we introduce a novel mechanism called \underline{\textbf{D}}irect Preference Learning with Only \underline{\textbf{S}}elf-Generated \underline{\textbf{T}}ests and \underline{\textbf{C}}ode (DSTC), detailed in Algorithm \ref{alg1} and illustrated in Fig. \ref{fig:illus}. DSTC begins with prompting the model to generate multiple code snippets and tests for each given instruction. Given self-generated code snippets and tests, DSTC comprises two key components: minimax selection and code-test concatenation. Intuitively, DSTC extracts more reliable preference pairs from self-generated code snippets and tests while mitigating the misleading effects of incorrect tests or code.
Using the preference dataset constructed by DSTC, direct preference learning algorithms such as DPO and KTO can be effectively employed to train code LMs.

\paragraph{Preprocess: Generating code snippets and tests.}(Lines \ref{line:0_start}-\ref{line:0_end} in Algorithm \ref{alg1}) In the preprocessing phase of DSTC, we prompt the model to generate $J$ pairs of code snippets $\{y_i^j\}_{j=1}^J$ and tests $\{z_i^j\}_{j=1}^J$ for each instruction $x_i$ in the instruction set.
Here, we adapt the prompt template\footnote{\url{https://github.com/bigcode-project/starcoder2-self-align}} from the \textit{oss-instruct-sd2-50k} dataset \citep{wei2024selfcodealign} to generate high-quality code and tests. We show the details of the prompt in Appendix \ref{app:prompt}. To mitigate errors caused by mismatched entry points (e.g., function or class names) during the evaluation from a compiler, such as when assessing the correctness of a code snippet $y_i^{j_1}$ with a test $z_i^{j_2}$ where $j_1 \neq j_2$, we include entry points in the prompts. These entry points ensure the consistency

\begin{algorithm}[H]
	\caption{Direct Preference Learning with Only Self-Generated Tests and Code (DSTC)}
	\label{alg1}
	\begin{algorithmic}[1]
	\STATE \textbf{Input}: Instruction set $\mathcal{I}=\{x_i\}_{i=1}^I$, sampling time $J$, model $\pi$, pre-defined prompt \texttt{prompt} for concatenation, and a compiler for returning binary execution feedback.
 \STATE Initialize the preference dataset $\mathcal{D}=\emptyset$.
 \label{line:0_start}
 \FOR{$i=1,\ldots, I$}
\FOR{$j=1,\ldots, J$}
\STATE Prompt the model $\pi$ to sample the code $y_{i}^{j}$ and test $z_i^j$ given the instruction $x_i$.  
 \ENDFOR \label{line:0_end}
\FOR{$j=1,\ldots, J$}\label{line:1_start}
\FOR{$k=1,\ldots, J$}
\STATE Concatenate the code $y_i^j$ and test $z_i^k$ to execute in the sandbox and receive the binary feedback $r_{ijk}\in\{0,1\}$.
 \ENDFOR  
 \STATE Select $j^\prime = \argmax_j \sum_{k=1}^J r_{ijk} $. \hfill(\textit{Select the chosen code}) \label{l:1}
 \STATE Select $k^\prime = \argmin_k \sum_{j=1}^J r_{ijk} \quad\text{ s.t. } r_{ij^\prime k} = 1$. \hfill(\textit{Select the chosen test}) \label{l:2}
  \STATE Select $k^\dagger = \argmax_k \sum_{j=1}^J r_{ijk} \quad\text{ s.t. } \sum_{j=1}^J r_{ijk} < J $. \hfill(\textit{Select the rejected test}) \label{l:3}
  \STATE Select $j^\dagger = \argmin_j \sum_{k=1}^J r_{ijk} 
\quad\text{ s.t. } r_{ij k^\dagger} = 0$. \hfill(\textit{Select the rejected code}) \label{l:4}
  \ENDFOR\label{line:1_end}
  \STATE \textbf{Option I (DPO)}: \label{line:2_start}
            \IF{$j^\dagger$, $k^\dagger$, and $k^\prime$ are all not \texttt{None} }
            \STATE Concatenate code $y_i^{j^\prime}$, pre-defined prompt \texttt{prompt}, and test $z_i^{k^\prime}$ to get the chosen response $a_{i}^{\text{chosen}} = \texttt{Concat}(y_i^{j^\prime}, \texttt{prompt},z_i^{k^\prime})$.
             \STATE Concatenate code $y_i^{j^\dagger}$, pre-defined prompt \texttt{prompt}, and test $z_i^{k^\dagger}$ to get the rejected response $a_{i}^{\text{rejected}} = \texttt{Concat}(y_i^{j^\dagger}, \texttt{prompt},z_i^{k^\dagger})$.
            \STATE Add $(x_i, a_{i}^{\text{chosen}}, a_{i}^{\text{rejected}})$ in $\mathcal{D}$.
            \ENDIF
    \STATE \textbf{Option II (KTO)}: 
            \IF{$k^\prime$ is not \texttt{None} }
            \STATE Concatenate code $y_i^{j^\prime}$, pre-defined prompt \texttt{prompt}, and test $z_i^{k^\prime}$ to get the chosen response $a_{i}^{\text{chosen}} = \texttt{Concat}(y_i^{j^\prime}, \texttt{prompt},z_i^{k^\prime})$.
            \STATE Add $(x_i, a_{i}^{\text{chosen}}, 1)$ in $\mathcal{D}$.
            \IF{$j^\dagger$ and $k^\dagger$ are both not \texttt{None}}
            \STATE Concatenate code $y_i^{j^\dagger}$, pre-defined prompt \texttt{prompt}, and test $z_i^{k^\dagger}$ to get the rejected response $a_{i}^{\text{rejected}} = \texttt{Concat}(y_i^{j^\dagger}, \texttt{prompt},z_i^{k^\dagger})$.
            \STATE Add $(x_i, a_{i}^{\text{rejected}}, 0)$
            \ENDIF
            \ENDIF
   \ENDFOR\label{line:2_end}
    \STATE \textbf{Output}: Preference dataset $\cD$.
	\end{algorithmic}
\end{algorithm}
\noindent between the code snippets and tests during evaluation. After the preprocessing phase, the entry points are removed to prevent potential information leakage that could lead to overfitting during training.
\paragraph{Component I: Minimax selection.}(Lines \ref{line:1_start}-\ref{line:1_end} in Algorithm \ref{alg1}) For each self-generated code snippet $y_i^j$ and test $z_i^j$, we concatenate them in pairs and record their binary execution feedback $r_{ijk}\in\{0,1\}$ from a compiler.  This feedback determines which code snippets and tests are chosen or rejected based on the following criteria (if an optimization problem is infeasible, the solution defaults to \texttt{None}):
\begin{itemize}
    \item \textbf{Selection of the chosen code snippet.} (Line \ref{l:1} in Algorithm \ref{alg1})  For the chosen code, we wish to select the code snippet that is the \textbf{most likely to be correct}. By measuring the correctness via the number of passing tests, we formulate the selection of the chosen code snippet $y_i^{j^\prime}$ as the following optimization
    \begin{align}
        j^\prime = \argmax_j \sum_{k=1}^J r_{ijk}.\label{eq:1}
    \end{align}

    \item \textbf{Selection of the chosen test.} (Line \ref{l:2} in Algorithm \ref{alg1})
    For the chosen test, we wish to select the test that is the \textbf{most difficult to pass} that can be evaluated by the number of passing code snippets. We formulate the selection of the chosen test $z_i^{k^\prime}$ as the following constraint optimization
\begin{align}
    k^\prime = \argmin_k \sum_{j=1}^J r_{ijk} \quad\text{ s.t. } r_{ij^\prime k} = 1, \label{eq:2}
\end{align} 
 where the constraint ensures that the chosen code snippet passes the chosen test.
 
    \item \textbf{Selection of the rejected test.} (Line \ref{l:3} in Algorithm \ref{alg1}) 
For the rejected test, we wish to select the test that is the \textbf{easiest to pass}. In contrast to \eqref{eq:2}, we formulate the selection of the rejected test $z_i^{k^\dagger}$ as the following constraint optimization
\begin{align}
    k^\dagger = \argmax_k \sum_{j=1}^J r_{ijk} \quad\text{ s.t. } \sum_{j=1}^J r_{ijk} < J,
\end{align}
where the constraint ensures that at least one code snippet fails the rejected test.
        \item \textbf{Selection of the rejected code snippet.} (Line \ref{l:4} in Algorithm \ref{alg1})
        For the rejected code snippet, we wish to select the code snippet that is the \textbf{least likely to be correct}. In contrast to \eqref{eq:1}, we formulate the selection of the rejected code snippet $y_i^{k^\dagger}$ as the following constraint optimization
        \begin{align}
            j^\dagger = \argmin_j \sum_{k=1}^J r_{ijk} 
\quad\text{ s.t. } r_{ij k^\dagger} = 0,
        \end{align}
         where the constraint ensures that the failed code snippet fails the rejected test.
\end{itemize}

\paragraph{Component II: Concatenation between the code snippet and test.}(Lines \ref{line:2_start}-\ref{line:2_end} in Algorithm \ref{alg1})
After the phase of the minimax selection in Algorithm \ref{alg1}, we concatenate the chosen (resp. rejected) code snippet and test as the chosen (resp. rejected) response with the following prompt template for concatenation:
\begin{tcolorbox}

[\textbf{selected code snippet}]

The provided code should satisfy the following assertions:

[\textbf{selected test}]

\end{tcolorbox}
The idea of this concatenation is to form a preference pair with a reliable order brought by the execution feedback. Depending on which specific preference learning algorithm we use, we consider different manners to construct the preference dataset. For DPO, the preference dataset involves chosen-rejected pairs, hence we filter out the data without both feasible chosen and rejected responses generated via the minimax selection. For KTO, we only filter out the data without feasible chosen responses, as KTO preference dataset permits an imbalance between the number of chosen and rejected responses.
With the preference dataset generated by DSTC, we incorporate the corresponding direct preference learning algorithm to train the code LMs.

\paragraph{DSTC generates more reliable preference pairs.}

\begin{table}[H]
    \centering
   \resizebox{1\textwidth}{!}
    {
    \begin{tabular}{  c  c c  c}
    \toprule
    \multirow{2}{*}{Method}  & \multicolumn{2}{c}{Accuracy} & \multirow{2}{*}{Chosen Code $>$ Rejected Code (\%)}\\
    \cline{2-3} 
         & Chosen Code (\%)  & Rejected Code (\%) \\
        \midrule
        Baseline &
45.1 &
23.1 &
26.2 \\
DSTC & 
51.9 &
16.7 &
36.9 
\\
    \bottomrule
    \end{tabular}%
    }
    \caption{Results on the preference dataset generated by \textit{Starcoder2-15B-instruct}, where Chosen Code $>$ Rejected Code means that the chosen code snippet passes the test but the rejected code snippet fails. We use \textit{DeepSeek-Coder-V2} with 236 billion parameters to generate the tests for evaluation.}
    \label{tab:analysis}
\end{table}
 In Table \ref{tab:analysis}, we show that our proposed method DSTC can improve the code quality and enlarge the quality gap between the chosen code and the rejected code, making the preference pair more reliable. Since there are still 36.9\% pairs of chosen code and rejected code in DTSC that they both fail or pass the corresponding tests, it outlines the necessity to incorporate the code-test concatenation to generate a more reliable preference dataset. 

\section{Experiments}
In this section, we conduct extensive experiments to show the effectiveness of our proposed DSTC in improving LMs' code generation ability. 
\subsection{Experimental Setup}
We implement DSTC on two both open-resource fine-tuned models \textit{StarCoder2-15b-Instruct}\footnote{\url{https://huggingface.co/bigcode/starcoder2-15b-instruct-v0.1}} and \textit{Deepseek-coder-33b-instruct}\footnote{\url{https://huggingface.co/deepseek-ai/deepseek-coder-33b-instruct}}, respectively. 
We use the instructions from the \textit{oss-instruct-sd2-50k} dataset\footnote{\url{https://huggingface.co/datasets/bigcode/self-oss-instruct-sc2-exec-filter-50k}} to form the instruction set, where all the instructions are generated and filtered by \textit{Starcoder2-15b} given seed functions. Specifically, \textit{StarCoder2-15b-Instruct} is supervised fine-tuned (SFT) on the \textit{oss-instruct-sd2-50k} dataset without further RLHF training. 
We use the implementation of DPO and KTO in the  OpenRLHF codebase\footnote{\url{https://github.com/OpenRLHF/OpenRLHF}} \citep{hu2024openrlhf} as our training framework. The detailed training configurations are provided in Appendix  \ref{app:detail_train}.
\paragraph{Benchmarks.}
We evaluate the LLM coding generation ability by five benchmarks: HumanEval Base \citep{chen2021evaluating}, HumanEval Plus \citep{liu2024your}, Mostly Basic Python Problems (MBPP) Base \citep{austin2021program}, MBPP Plus \citep{liu2024your}, and  BigCodeBench (BCB) \citep{zhuo2024bigcodebench}. In particular, BCB consists of two splits: \textit{instruct split} and \textit{complete split}, where \textit{instruct split} only involves NLP instructions  and \textit{complete split} uses structured docstring in
prompts.
For each split, \textit{hard subset} in BCB contains the most challenging and user-centric fraction of the \textit{full set}. During the evaluation of all these benchmarks, we use greedy decoding in the code generation and all results are reported as the pass@1 score. Here, the pass@1 score measures the accuracy of code LMs on their initial attempt at generating the correct code.

\subsection{DSTC Enhances Code LMs Across Diverse Benchmarks}
The results in Table \ref{tab:benchmark1} demonstrate that DSTC consistently enhances the performance of code LMs, surpassing baseline models by a stable margin across all benchmarks when combined with either DPO or KTO. Notably, DSTC achieves improvements in the pass@1 score without requiring additional resources, such as canonical solutions or synthetic tests generated by ChatGPT. This highlights its potential for developing automated systems to further advance code LMs effectively.
\begin{table}[h]
    \centering
   \resizebox{1\textwidth}{!}
    {
    \begin{tabular}{    c c  cccc ccc}
    \toprule
    \multirow{2}{*}{Model}  & \multicolumn{2}{c}{HumanEval} & \multicolumn{2}{c}{MBPP}  & \multicolumn{2}{c}{BCB (Complete)} & \multicolumn{2}{c}{BCB (Instruct)}\\
    \cline{2-3} \cline{4-5}\cline{6-7}\cline{8-9}
         & Basic (\%)  & Plus (\%) &Basic (\%) & Plus  (\%) & Hard (\%)&
         Full (\%) & Hard (\%)& Full (\%)\\
        \midrule\midrule 
        Ref. & 72.0 &
62.2 & 75.2 & 60.9 & 
14.9 & 45.1&
12.2 &
37.6\\
\midrule 
DSTC + DPO & \textbf{74.4} &
\textbf{66.5} & 76.4 & 61.9 & 
\textbf{16.2} &\textbf{47.5}&
\textbf{12.8} &
37.9\\
DPO & 70.1 &
62.2 & \textbf{76.9} & \textbf{62.2} & 
14.2 & 46.6&
12.2 &
\textbf{38.1}\\
 \midrule 
DSTC + KTO & \textbf{73.8} &
\textbf{66.5} &
\textbf{75.9} &
\textbf{61.4} & 
\textbf{16.9} &\textbf{47.1} & 
\textbf{14.2} &

\textbf{38.9} \\ 
KTO &
62.8 &
54.3 &
74.9 &
61.2 &
15.5 & 46.6 &
8.1 &

38.1 
\\
    \bottomrule
    \end{tabular}%
    }
    \caption{Evaluation results of DSTC incorporated with DPO Dand KTO on \textit{Starcoder2-15b-instruct}.}
    \label{tab:benchmark1}
\end{table}

\subsection{Ablation Study}
\paragraph{DSTC can improve code LM with a larger size.}
Evaluation results in Table \ref{tab:benchmark2} showcase the effectiveness of DSTC when applied to the \textit{DeepseekCoder-33b-instruct}, a large-scale code language model with 33 billion parameters. These findings underscore the scalability of DSTC, demonstrating its ability to improve performance even in substantially larger models. 
\begin{table}[H]
    \centering
   \resizebox{1\textwidth}{!}
    {
    \begin{tabular}{    c c  cccc ccc}
    \toprule
    \multirow{2}{*}{Model}  & \multicolumn{2}{c}{HumanEval} & \multicolumn{2}{c}{MBPP}  & \multicolumn{2}{c}{BCB (Complete)} & \multicolumn{2}{c}{BCB (Instruct)}\\
    \cline{2-3} \cline{4-5}\cline{6-7}\cline{8-9}
         & Basic (\%)  & Plus (\%) &Basic (\%) & Plus  (\%) & Hard (\%)&
         Full (\%) & Hard (\%)& Full (\%)\\
        \midrule
        Ref. &
77.4 &
70.1 &
81.7 &
69.0 &
19.6 &50.9 &
14.9 &

\textbf{42.0} \\
DSTC & 
\textbf{79.9} &
\textbf{72.0} &
\textbf{82.5} &
\textbf{70.4} &
\textbf{22.3} &\textbf{51.6} &
\textbf{18.2} &

{41.0}
\\
    \bottomrule
    \end{tabular}%
    }
    \caption{Evaluation results of DSTC incorporated with DPO on \textit{DeepseekCoder-33b-instruct}.}
    \label{tab:benchmark2}
\end{table}
\paragraph{Both components in DSTC are necessary.} To study the effect of each component in DSTC, we perform an ablation study on DSTC and report the results in Table \ref{tab:benchmark3}, which shows that removing any single component leads to a measurable decline in performance. These results highlight the integral role of each component of DSTC in boosting the model's ability to generate accurate and reliable code.

\begin{table}[H]
    \centering
   \resizebox{1\textwidth}{!}
    {
    \begin{tabular}{    c c  cccc ccc}
    \toprule
    \multirow{2}{*}{Model}  & \multicolumn{2}{c}{HumanEval} & \multicolumn{2}{c}{MBPP}  & \multicolumn{2}{c}{BCB (Complete)} & \multicolumn{2}{c}{BCB (Instruct)}\\
    \cline{2-3} \cline{4-5}\cline{6-7}\cline{8-9}
         & Basic (\%)  & Plus (\%) &Basic (\%) & Plus  (\%) & Hard (\%)&
         Full (\%) & Hard (\%)& Full (\%)\\
        \midrule\midrule 
        Ref. & 72.0 &
62.2 & 75.2 & 60.9 & 
14.9 & 45.1&
12.2 &
37.6\\
\midrule 
DSTC & \textbf{74.4} &
\textbf{66.5} & 76.4 & 61.9 & 
\textbf{16.2} &
\textbf{47.5}& 
{12.8} &37.9\\
DSTC w/o Concatenation
 &72.0
&
65.2
&
\textbf{76.7} &
\textbf{62.2} &
15.5 & 47.0 &
12.8 &

\textbf{38.4}\\
DSTC w/o Minimax Selection &
71.3  &
62.2 &
75.4 &
61.4 &
15.5 & 46.5 &
\textbf{14.2} &

37.5
\\
    \bottomrule
    \end{tabular}%
    }
    \caption{Ablation study of DSTC incorporated with DPO on \textit{Starcoder2-15b-instruct}.}
    \label{tab:benchmark3}
\end{table}

\section{Conclusion}

In this paper, we propose a novel mechanism named DSTC to incorporate direct preference learning methods such as DPO and KTO to enhance the code generation capabilities of LMs using only self-generated code snippets and tests to create preference pairs, circumventing the need for costly annotated datasets. By employing a minimax selection process and concatenated test-code pairs, DSTC mitigates the impact of low-quality self-generated tests, enabling stable performance gains across multiple coding benchmarks. Applied with direct preference learning methods like DPO and KTO, DSTC consistently improves pass@1 scores, showing both scalability and generalizability for models of different sizes. This work highlights DSTC’s potential to advance coding accuracy autonomously, marking a step forward in resource-efficient, high-performance code LM training. Future directions may explore additional mechanisms for refining self-generated tests and expanding DSTC’s applicability across broader coding tasks.

\bibliographystyle{ims}

\bibliography{main}

\begin{thebibliography}{47}
\expandafter\ifx\csname natexlab\endcsname\relax\def\natexlab#1{#1}\fi
\expandafter\ifx\csname url\endcsname\relax
  \def\url#1{\texttt{#1}}\fi
\expandafter\ifx\csname urlprefix\endcsname\relax\def\urlprefix{}\fi

\bibitem[{Achiam et~al.(2023)Achiam, Adler, Agarwal, Ahmad, Akkaya, Aleman, Almeida, Altenschmidt, Altman, Anadkat et~al.}]{achiam2023gpt}
\text{Achiam, J.}, \text{Adler, S.}, \text{Agarwal, S.}, \text{Ahmad, L.}, \text{Akkaya, I.}, \text{Aleman, F.~L.}, \text{Almeida, D.}, \text{Altenschmidt, J.}, \text{Altman, S.}, \text{Anadkat, S.} \text{et~al.} (2023).
\newblock Gpt-4 technical report.
\newblock \textit{arXiv preprint arXiv:2303.08774}.

\bibitem[{Ahn et~al.(2024)Ahn, Verma, Lou, Liu, Zhang and Yin}]{ahn2024large}
\text{Ahn, J.}, \text{Verma, R.}, \text{Lou, R.}, \text{Liu, D.}, \text{Zhang, R.} and \text{Yin, W.} (2024).
\newblock Large language models for mathematical reasoning: Progresses and challenges.
\newblock \textit{arXiv preprint arXiv:2402.00157}.

\bibitem[{Anthropic(2023)}]{claude}
\text{Anthropic} (2023).
\newblock Introducing claude.
\newblock \textit{https://www.anthropic.com/news/introducing-claude}.

\bibitem[{Austin et~al.(2021)Austin, Odena, Nye, Bosma, Michalewski, Dohan, Jiang, Cai, Terry, Le et~al.}]{austin2021program}
\text{Austin, J.}, \text{Odena, A.}, \text{Nye, M.}, \text{Bosma, M.}, \text{Michalewski, H.}, \text{Dohan, D.}, \text{Jiang, E.}, \text{Cai, C.}, \text{Terry, M.}, \text{Le, Q.} \text{et~al.} (2021).
\newblock Program synthesis with large language models.
\newblock \textit{arXiv preprint arXiv:2108.07732}.

\bibitem[{Azar et~al.(2023)Azar, Rowland, Piot, Guo, Calandriello, Valko and Munos}]{azar2023general}
\text{Azar, M.~G.}, \text{Rowland, M.}, \text{Piot, B.}, \text{Guo, D.}, \text{Calandriello, D.}, \text{Valko, M.} and \text{Munos, R.} (2023).
\newblock A general theoretical paradigm to understand learning from human preferences.
\newblock \textit{arXiv preprint arXiv:2310.12036}.

\bibitem[{Bradley and Terry(1952)}]{bradley1952rank}
\text{Bradley, R.~A.} and \text{Terry, M.~E.} (1952).
\newblock Rank analysis of incomplete block designs: I. the method of paired comparisons.
\newblock \textit{Biometrika}, \textbf{39} 324--345.

\bibitem[{Chen et~al.(2021)Chen, Tworek, Jun, Yuan, Pinto, Kaplan, Edwards, Burda, Joseph, Brockman et~al.}]{chen2021evaluating}
\text{Chen, M.}, \text{Tworek, J.}, \text{Jun, H.}, \text{Yuan, Q.}, \text{Pinto, H. P. D.~O.}, \text{Kaplan, J.}, \text{Edwards, H.}, \text{Burda, Y.}, \text{Joseph, N.}, \text{Brockman, G.} \text{et~al.} (2021).
\newblock Evaluating large language models trained on code.
\newblock \textit{arXiv preprint arXiv:2107.03374}.

\bibitem[{Dou et~al.(2024)Dou, Liu, Jia, Xiong, Zhou, Shan, Huang, Shen, Fan, Xi et~al.}]{dou2024stepcoder}
\text{Dou, S.}, \text{Liu, Y.}, \text{Jia, H.}, \text{Xiong, L.}, \text{Zhou, E.}, \text{Shan, J.}, \text{Huang, C.}, \text{Shen, W.}, \text{Fan, X.}, \text{Xi, Z.} \text{et~al.} (2024).
\newblock Stepcoder: Improve code generation with reinforcement learning from compiler feedback.
\newblock \textit{arXiv preprint arXiv:2402.01391}.

\bibitem[{Engstrom et~al.(2020)Engstrom, Ilyas, Santurkar, Tsipras, Janoos, Rudolph and Madry}]{engstrom2020implementation}
\text{Engstrom, L.}, \text{Ilyas, A.}, \text{Santurkar, S.}, \text{Tsipras, D.}, \text{Janoos, F.}, \text{Rudolph, L.} and \text{Madry, A.} (2020).
\newblock Implementation matters in deep policy gradients: A case study on ppo and trpo.
\newblock \textit{arXiv preprint arXiv:2005.12729}.

\bibitem[{Ethayarajh et~al.(2024)Ethayarajh, Xu, Muennighoff, Jurafsky and Kiela}]{ethayarajh2024kto}
\text{Ethayarajh, K.}, \text{Xu, W.}, \text{Muennighoff, N.}, \text{Jurafsky, D.} and \text{Kiela, D.} (2024).
\newblock Kto: Model alignment as prospect theoretic optimization.
\newblock \textit{arXiv preprint arXiv:2402.01306}.

\bibitem[{Gee et~al.(2024)Gee, Gritta, Lampouras and Iacobacci}]{gee2024code}
\text{Gee, L.}, \text{Gritta, M.}, \text{Lampouras, G.} and \text{Iacobacci, I.} (2024).
\newblock Code-optimise: Self-generated preference data for correctness and efficiency.
\newblock \textit{arXiv preprint arXiv:2406.12502}.

\bibitem[{Hong et~al.(2024)Hong, Lee and Thorne}]{hong2024orpo}
\text{Hong, J.}, \text{Lee, N.} and \text{Thorne, J.} (2024).
\newblock Orpo: Monolithic preference optimization without reference model.
\newblock \textit{arXiv preprint arXiv:2403.07691}.

\bibitem[{Hu et~al.(2024)Hu, Wu, Wang, Xianyu, Zhang and Cao}]{hu2024openrlhf}
\text{Hu, J.}, \text{Wu, X.}, \text{Wang, W.}, \text{Xianyu}, \text{Zhang, D.} and \text{Cao, Y.} (2024).
\newblock Openrlhf: An easy-to-use, scalable and high-performance rlhf framework.
\newblock \textit{arXiv preprint arXiv:2405.11143}.

\bibitem[{Hui et~al.(2024)Hui, Yang, Cui, Yang, Liu, Zhang, Liu, Zhang, Yu, Dang et~al.}]{hui2024qwen2}
\text{Hui, B.}, \text{Yang, J.}, \text{Cui, Z.}, \text{Yang, J.}, \text{Liu, D.}, \text{Zhang, L.}, \text{Liu, T.}, \text{Zhang, J.}, \text{Yu, B.}, \text{Dang, K.} \text{et~al.} (2024).
\newblock Qwen2. 5-coder technical report.
\newblock \textit{arXiv preprint arXiv:2409.12186}.

\bibitem[{Ji et~al.(2024)Ji, Lu, Niu, Ke, Wang, Zhu, Tang and Huang}]{ji2024towards}
\text{Ji, H.}, \text{Lu, C.}, \text{Niu, Y.}, \text{Ke, P.}, \text{Wang, H.}, \text{Zhu, J.}, \text{Tang, J.} and \text{Huang, M.} (2024).
\newblock Towards efficient and exact optimization of language model alignment.
\newblock \textit{arXiv preprint arXiv:2402.00856}.

\bibitem[{Jiang et~al.(2024)Jiang, Wang, Shen, Kim and Kim}]{jiang2024survey}
\text{Jiang, J.}, \text{Wang, F.}, \text{Shen, J.}, \text{Kim, S.} and \text{Kim, S.} (2024).
\newblock A survey on large language models for code generation.
\newblock \textit{arXiv preprint arXiv:2406.00515}.

\bibitem[{Le et~al.(2022)Le, Wang, Gotmare, Savarese and Hoi}]{le2022coderl}
\text{Le, H.}, \text{Wang, Y.}, \text{Gotmare, A.~D.}, \text{Savarese, S.} and \text{Hoi, S. C.~H.} (2022).
\newblock Coderl: Mastering code generation through pretrained models and deep reinforcement learning.
\newblock \textit{Advances in Neural Information Processing Systems}, \textbf{35} 21314--21328.

\bibitem[{Liang et~al.(2024)Liang, Chen, Wang, Wu, Fu, Shi, Wu and Ye}]{liang2024robust}
\text{Liang, X.}, \text{Chen, C.}, \text{Wang, J.}, \text{Wu, Y.}, \text{Fu, Z.}, \text{Shi, Z.}, \text{Wu, F.} and \text{Ye, J.} (2024).
\newblock Robust preference optimization with provable noise tolerance for llms.
\newblock \textit{arXiv preprint arXiv:2404.04102}.

\bibitem[{Liu et~al.(2024{\natexlab{a}})Liu, Xia, Wang and Zhang}]{liu2024your}
\text{Liu, J.}, \text{Xia, C.~S.}, \text{Wang, Y.} and \text{Zhang, L.} (2024{\natexlab{a}}).
\newblock Is your code generated by chatgpt really correct? rigorous evaluation of large language models for code generation.
\newblock \textit{Advances in Neural Information Processing Systems}, \textbf{36}.

\bibitem[{Liu et~al.(2023{\natexlab{a}})Liu, Zhu, Xiao, Fu, Han, Yang and Ye}]{liu2023rltf}
\text{Liu, J.}, \text{Zhu, Y.}, \text{Xiao, K.}, \text{Fu, Q.}, \text{Han, X.}, \text{Yang, W.} and \text{Ye, D.} (2023{\natexlab{a}}).
\newblock Rltf: Reinforcement learning from unit test feedback.
\newblock \textit{arXiv preprint arXiv:2307.04349}.

\bibitem[{Liu et~al.(2023{\natexlab{b}})Liu, Zhao, Joshi, Khalman, Saleh, Liu and Liu}]{liu2023statistical}
\text{Liu, T.}, \text{Zhao, Y.}, \text{Joshi, R.}, \text{Khalman, M.}, \text{Saleh, M.}, \text{Liu, P.~J.} and \text{Liu, J.} (2023{\natexlab{b}}).
\newblock Statistical rejection sampling improves preference optimization.
\newblock \textit{arXiv preprint arXiv:2309.06657}.

\bibitem[{Liu et~al.(2024{\natexlab{b}})Liu, Lu, Zhang, Liu, Guo, Yang, Blanchet and Wang}]{liu2024provably}
\text{Liu, Z.}, \text{Lu, M.}, \text{Zhang, S.}, \text{Liu, B.}, \text{Guo, H.}, \text{Yang, Y.}, \text{Blanchet, J.} and \text{Wang, Z.} (2024{\natexlab{b}}).
\newblock Provably mitigating overoptimization in rlhf: Your sft loss is implicitly an adversarial regularizer.
\newblock \textit{arXiv preprint arXiv:2405.16436}.

\bibitem[{Lozhkov et~al.(2024)Lozhkov, Li, Allal, Cassano, Lamy-Poirier, Tazi, Tang, Pykhtar, Liu, Wei et~al.}]{lozhkov2024starcoder}
\text{Lozhkov, A.}, \text{Li, R.}, \text{Allal, L.~B.}, \text{Cassano, F.}, \text{Lamy-Poirier, J.}, \text{Tazi, N.}, \text{Tang, A.}, \text{Pykhtar, D.}, \text{Liu, J.}, \text{Wei, Y.} \text{et~al.} (2024).
\newblock Starcoder 2 and the stack v2: The next generation.
\newblock \textit{arXiv preprint arXiv:2402.19173}.

\bibitem[{Luo et~al.(2023)Luo, Xu, Zhao, Sun, Geng, Hu, Tao, Ma, Lin and Jiang}]{luo2023wizardcoder}
\text{Luo, Z.}, \text{Xu, C.}, \text{Zhao, P.}, \text{Sun, Q.}, \text{Geng, X.}, \text{Hu, W.}, \text{Tao, C.}, \text{Ma, J.}, \text{Lin, Q.} and \text{Jiang, D.} (2023).
\newblock Wizardcoder: Empowering code large language models with evol-instruct.
\newblock \textit{arXiv preprint arXiv:2306.08568}.

\bibitem[{Pal et~al.(2024)Pal, Karkhanis, Dooley, Roberts, Naidu and White}]{pal2024smaug}
\text{Pal, A.}, \text{Karkhanis, D.}, \text{Dooley, S.}, \text{Roberts, M.}, \text{Naidu, S.} and \text{White, C.} (2024).
\newblock Smaug: Fixing failure modes of preference optimisation with dpo-positive.
\newblock \textit{arXiv preprint arXiv:2402.13228}.

\bibitem[{Rafailov et~al.(2023)Rafailov, Sharma, Mitchell, Manning, Ermon and Finn}]{rafailov2024direct}
\text{Rafailov, R.}, \text{Sharma, A.}, \text{Mitchell, E.}, \text{Manning, C.~D.}, \text{Ermon, S.} and \text{Finn, C.} (2023).
\newblock Direct preference optimization: Your language model is secretly a reward model.
\newblock \textit{Advances in Neural Information Processing Systems}, \textbf{36}.

\bibitem[{Rosset et~al.(2024)Rosset, Cheng, Mitra, Santacroce, Awadallah and Xie}]{rosset2024direct}
\text{Rosset, C.}, \text{Cheng, C.-A.}, \text{Mitra, A.}, \text{Santacroce, M.}, \text{Awadallah, A.} and \text{Xie, T.} (2024).
\newblock Direct nash optimization: Teaching language models to self-improve with general preferences.
\newblock \textit{arXiv preprint arXiv:2404.03715}.

\bibitem[{Roziere et~al.(2023)Roziere, Gehring, Gloeckle, Sootla, Gat, Tan, Adi, Liu, Sauvestre, Remez et~al.}]{roziere2023code}
\text{Roziere, B.}, \text{Gehring, J.}, \text{Gloeckle, F.}, \text{Sootla, S.}, \text{Gat, I.}, \text{Tan, X.~E.}, \text{Adi, Y.}, \text{Liu, J.}, \text{Sauvestre, R.}, \text{Remez, T.} \text{et~al.} (2023).
\newblock Code llama: Open foundation models for code.
\newblock \textit{arXiv preprint arXiv:2308.12950}.

\bibitem[{Schulman et~al.(2017)Schulman, Wolski, Dhariwal, Radford and Klimov}]{schulman2017proximal}
\text{Schulman, J.}, \text{Wolski, F.}, \text{Dhariwal, P.}, \text{Radford, A.} and \text{Klimov, O.} (2017).
\newblock Proximal policy optimization algorithms.
\newblock \textit{arXiv preprint arXiv:1707.06347}.

\bibitem[{Shen and Zhang(2024)}]{shen2024policy}
\text{Shen, W.} and \text{Zhang, C.} (2024).
\newblock Policy filtration in rlhf to fine-tune llm for code generation.
\newblock \textit{arXiv preprint arXiv:2409.06957}.

\bibitem[{Shojaee et~al.(2023)Shojaee, Jain, Tipirneni and Reddy}]{shojaee2023execution}
\text{Shojaee, P.}, \text{Jain, A.}, \text{Tipirneni, S.} and \text{Reddy, C.~K.} (2023).
\newblock Execution-based code generation using deep reinforcement learning.
\newblock \textit{arXiv preprint arXiv:2301.13816}.

\bibitem[{Tajwar et~al.(2024)Tajwar, Singh, Sharma, Rafailov, Schneider, Xie, Ermon, Finn and Kumar}]{tajwar2024preference}
\text{Tajwar, F.}, \text{Singh, A.}, \text{Sharma, A.}, \text{Rafailov, R.}, \text{Schneider, J.}, \text{Xie, T.}, \text{Ermon, S.}, \text{Finn, C.} and \text{Kumar, A.} (2024).
\newblock Preference fine-tuning of llms should leverage suboptimal, on-policy data.
\newblock \textit{arXiv preprint arXiv:2404.14367}.

\bibitem[{Tang et~al.(2024)Tang, Guo, Zheng, Calandriello, Munos, Rowland, Richemond, Valko, Pires and Piot}]{tang2024generalized}
\text{Tang, Y.}, \text{Guo, Z.~D.}, \text{Zheng, Z.}, \text{Calandriello, D.}, \text{Munos, R.}, \text{Rowland, M.}, \text{Richemond, P.~H.}, \text{Valko, M.}, \text{Pires, B.~{\'A}.} and \text{Piot, B.} (2024).
\newblock Generalized preference optimization: A unified approach to offline alignment.
\newblock \textit{arXiv preprint arXiv:2402.05749}.

\bibitem[{Team et~al.(2023)Team, Anil, Borgeaud, Wu, Alayrac, Yu, Soricut, Schalkwyk, Dai, Hauth et~al.}]{team2023gemini}
\text{Team, G.}, \text{Anil, R.}, \text{Borgeaud, S.}, \text{Wu, Y.}, \text{Alayrac, J.-B.}, \text{Yu, J.}, \text{Soricut, R.}, \text{Schalkwyk, J.}, \text{Dai, A.~M.}, \text{Hauth, A.} \text{et~al.} (2023).
\newblock Gemini: a family of highly capable multimodal models.
\newblock \textit{arXiv preprint arXiv:2312.11805}.

\bibitem[{Wei et~al.(2024{\natexlab{a}})Wei, Cassano, Liu, Ding, Jain, Mueller, de~Vries, Werra, Guha and ZHANG}]{wei2024selfcodealign}
\text{Wei, Y.}, \text{Cassano, F.}, \text{Liu, J.}, \text{Ding, Y.}, \text{Jain, N.}, \text{Mueller, Z.}, \text{de~Vries, H.}, \text{Werra, L.~V.}, \text{Guha, A.} and \text{ZHANG, L.} (2024{\natexlab{a}}).
\newblock Selfcodealign: Self-alignment for code generation.
\newblock In \textit{The Thirty-eighth Annual Conference on Neural Information Processing Systems}.
\newline\urlprefix\url{https://openreview.net/forum?id=xXRnUU7xTL}

\bibitem[{Wei et~al.(2024{\natexlab{b}})Wei, Wang, Liu, Ding and Zhang}]{wei2024magicoder}
\text{Wei, Y.}, \text{Wang, Z.}, \text{Liu, J.}, \text{Ding, Y.} and \text{Zhang, L.} (2024{\natexlab{b}}).
\newblock Magicoder: Empowering code generation with oss-instruct.
\newblock In \textit{Forty-first International Conference on Machine Learning}.

\bibitem[{Wu et~al.(2024)Wu, Sun, Yuan, Ji, Yang and Gu}]{wu2024self}
\text{Wu, Y.}, \text{Sun, Z.}, \text{Yuan, H.}, \text{Ji, K.}, \text{Yang, Y.} and \text{Gu, Q.} (2024).
\newblock Self-play preference optimization for language model alignment.
\newblock \textit{arXiv preprint arXiv:2405.00675}.

\bibitem[{Xiong et~al.(2023)Xiong, Dong, Ye, Zhong, Jiang and Zhang}]{xiong2023gibbs}
\text{Xiong, W.}, \text{Dong, H.}, \text{Ye, C.}, \text{Zhong, H.}, \text{Jiang, N.} and \text{Zhang, T.} (2023).
\newblock Gibbs sampling from human feedback: A provable kl-constrained framework for rlhf.
\newblock \textit{arXiv preprint arXiv:2312.11456}.

\bibitem[{Xu et~al.(2024)Xu, Liu, Liu, Yan, Wang, Zhang and He}]{xu2024just}
\text{Xu, R.}, \text{Liu, Z.}, \text{Liu, Y.}, \text{Yan, S.}, \text{Wang, Z.}, \text{Zhang, Z.} and \text{He, X.} (2024).
\newblock Just say what you want: Only-prompting self-rewarding online preference optimization.
\newblock \textit{arXiv preprint arXiv:2409.17534}.

\bibitem[{Ye et~al.(2024)Ye, Xiong, Zhang, Jiang and Zhang}]{ye2024theoretical}
\text{Ye, C.}, \text{Xiong, W.}, \text{Zhang, Y.}, \text{Jiang, N.} and \text{Zhang, T.} (2024).
\newblock A theoretical analysis of nash learning from human feedback under general kl-regularized preference.
\newblock \textit{arXiv preprint arXiv:2402.07314}.

\bibitem[{Zhang et~al.(2024{\natexlab{a}})Zhang, Diao, Zou and Peng}]{zhang2024plum}
\text{Zhang, D.}, \text{Diao, S.}, \text{Zou, X.} and \text{Peng, H.} (2024{\natexlab{a}}).
\newblock Plum: Preference learning plus test cases yields better code language models.
\newblock \textit{arXiv preprint arXiv:2406.06887}.

\bibitem[{Zhang et~al.(2024{\natexlab{b}})Zhang, Lin, Bai and Mei}]{zhang2024negative}
\text{Zhang, R.}, \text{Lin, L.}, \text{Bai, Y.} and \text{Mei, S.} (2024{\natexlab{b}}).
\newblock Negative preference optimization: From catastrophic collapse to effective unlearning.
\newblock \textit{arXiv preprint arXiv:2404.05868}.

\bibitem[{Zhang et~al.(2024{\natexlab{c}})Zhang, Liu, Liu, Zhang, Yang, Liu, Chen, Sun and Wang}]{zhang2024reward}
\text{Zhang, S.}, \text{Liu, Z.}, \text{Liu, B.}, \text{Zhang, Y.}, \text{Yang, Y.}, \text{Liu, Y.}, \text{Chen, L.}, \text{Sun, T.} and \text{Wang, Z.} (2024{\natexlab{c}}).
\newblock Reward-augmented data enhances direct preference alignment of llms.
\newblock \textit{arXiv preprint arXiv:2410.08067}.

\bibitem[{Zhang et~al.(2024{\natexlab{d}})Zhang, Yu, Sharma, Zhong, Liu, Yang, Wang, Hassan and Wang}]{zhang2024self}
\text{Zhang, S.}, \text{Yu, D.}, \text{Sharma, H.}, \text{Zhong, H.}, \text{Liu, Z.}, \text{Yang, Z.}, \text{Wang, S.}, \text{Hassan, H.} and \text{Wang, Z.} (2024{\natexlab{d}}).
\newblock Self-exploring language models: Active preference elicitation for online alignment.
\newblock \textit{arXiv preprint arXiv:2405.19332}.

\bibitem[{Zhao et~al.(2023)Zhao, Joshi, Liu, Khalman, Saleh and Liu}]{zhao2023slic}
\text{Zhao, Y.}, \text{Joshi, R.}, \text{Liu, T.}, \text{Khalman, M.}, \text{Saleh, M.} and \text{Liu, P.~J.} (2023).
\newblock Slic-hf: Sequence likelihood calibration with human feedback.
\newblock \textit{arXiv preprint arXiv:2305.10425}.

\bibitem[{Zhu et~al.(2024)Zhu, Guo, Shao, Yang, Wang, Xu, Wu, Li, Gao, Ma et~al.}]{zhu2024deepseek}
\text{Zhu, Q.}, \text{Guo, D.}, \text{Shao, Z.}, \text{Yang, D.}, \text{Wang, P.}, \text{Xu, R.}, \text{Wu, Y.}, \text{Li, Y.}, \text{Gao, H.}, \text{Ma, S.} \text{et~al.} (2024).
\newblock Deepseek-coder-v2: Breaking the barrier of closed-source models in code intelligence.
\newblock \textit{arXiv preprint arXiv:2406.11931}.

\bibitem[{Zhuo et~al.(2024)Zhuo, Vu, Chim, Hu, Yu, Widyasari, Yusuf, Zhan, He, Paul et~al.}]{zhuo2024bigcodebench}
\text{Zhuo, T.~Y.}, \text{Vu, M.~C.}, \text{Chim, J.}, \text{Hu, H.}, \text{Yu, W.}, \text{Widyasari, R.}, \text{Yusuf, I. N.~B.}, \text{Zhan, H.}, \text{He, J.}, \text{Paul, I.} \text{et~al.} (2024).
\newblock Bigcodebench: Benchmarking code generation with diverse function calls and complex instructions.
\newblock \textit{arXiv preprint arXiv:2406.15877}.

\end{thebibliography}

\newpage
\appendix
\section{Detailed Training Configurations}
\label{app:detail_train}
We train all the models with the OpenRLHF codebase \citep{hu2024openrlhf}.
We report the hyperparameters for the training in Table \ref{tab:hype}.

\begin{table}[H]
    \centering
     \resizebox{1\textwidth}{!}{
    \begin{tabular}{c |cc|c}
    \toprule 
        Base models & \multicolumn{2}{c}{\textit{Starcoder2-15b-instruct}} & \textit{Deepseek-coder-33b-instruct}\\
        \midrule 
        Preference learning & KTO & DPO & DPO\\
        \midrule
        Learning rate & 5.0e-7 & 5.0e-7& 5.0e-7\\
        Learning scheduler type& \texttt{cosine\_with\_min\_lr} &\texttt{cosine\_with\_min\_lr} &\texttt{cosine\_with\_min\_lr} \\
        Batch size &512 &256 & 256\\
        Micro batch size & 4 & 4 & 1 \\
        Training epoch & 1 &1 &1
        \\
        $\beta$ & 0.3 &0.2 &0.2\\
        Optimizer & \texttt{adamw} & \texttt{adamw} & \texttt{adamw}\\
        Precision & \texttt{bfloat16} & \texttt{bfloat16} & \texttt{bfloat16} \\
        Warmup ratio & 0.05 &0.05 &0.05
        \\
         \bottomrule
    \end{tabular}
    }
    \vspace{1mm}
    \caption{Training configurations for all the models in this paper.} 
    \label{tab:hype}
\end{table}

\section{Prompts for Generating Code Snippets and Tests}
\label{app:prompt}
In this section, we give details of the prompts used to generate code snippets and tests. We show an example of the full prompt as follows.
\vspace{0.25cm}
\begin{Verbatim}
Provide the best response to a given instruction. Follow the following steps to craft your response to fill Example 2:
1. First give a signal [Reasoning], then reason about the given instruction
2. First give a signal [Implementation], then provide a high-quality solution for the instruction.
3. First give a signal [Explanation], then offer a concise explanation.
4. First give a signal [Tests], write comprehensive and diverse tests carefully to verify the correctness your solution. Note that the tests should import all the required packages. 

## Example 1

### Instruction
Create a Python function to encode an integer into a sequence of encoded characters, where each character represents a 5-bit chunk of the integer. The ASCII value of each chunk character is formed by adding 63 to the 5-bit chunk's value. For all chunks except the last one, set the 6th bit to 1 to indicate continuation. Employ bitwise operations for this encoding process.
The main function name is `encode_integer_to_chars`.

### Response
We will organize our response by four parts: [Reasoning], [Implementation], [Explanation], and [Tests].

[Reasoning]
To encode an integer into a sequence of encoded characters based on the specified criteria, let's follow these steps:

1. Break down the integer into 5-bit chunks: Extract 5-bit chunks from the integer using using `& 31` (`0x1F`), because the binary representation of `31` (`11111`) has the last five bits set, which will effectively extract the 5 bits we are interested in. We start from the least significant bits (LSB).
2. Set continuation bit: For all chunks except the last one, set the 6th bit to 1 using `| 32` (`0x20`) because the binary representation of `32` (`100000`) has the 6th bit set to 1. This indicates that there's another continuation chunk following.
3. Shift right: Right-shift the value by 5 to process the next 5 bits.
4. Convert to characters: Use `ord` and `chr` to convert each chunk (now effectively 6 bits due to the continuation bit) into a character by adding 63 to its value. This ensures that the character is within a printable ASCII range.
5. Assembly: Assemble these characters into the final encoded string while maintaining the order of the chunks through `reversed`.

[Implementation]
Here's how you can implement this function:

```python
def encode_integer_to_chars(value: int) -> str:
    # Base case for 0
    if value == 0:
        return chr(63)

    encoded_chars = []
    while value > 0:
        # Extract the 5-bit chunk and add 63 to its value
        chunk = (value & 31) + 63
        value >>= 5
        # If there's still more of the value left, set the continuation bit
        if value > 0:
            chunk |= 32
        encoded_chars.append(chr(chunk))

    # Reverse the list of characters to start with the least significant chunk
    # and join them into a single string
    return ''.join(reversed(encoded_chars))
```

[Explanation]
This function processes the integer bit by bit in 5-bit chunks, applying bitwise operations to extract and manipulate these chunks, then converts each chunk into a character using the specified encoding scheme. This approach efficiently handles the encoding in a loop until all bits of the integer have been processed.

Note that this function is designed for non-negative integers and does not support negative integers by default because negative numbers are represented in two's complement form, where the most significant bit (MSB) indicates the sign. In Python, right-shifting (`>>`) a negative integer preserves the sign bit, effectively making the never executed due to the `value > 0` condition. If you change it to `value != 0`, it will then lead to an infinite loop if the same encoding logic is applied to negative integers without modifications.

[Tests]
You can validate this function with a few test cases to ensure correctness:

```python
assert encode_integer_to_chars(31) == chr(63 + 31)
assert encode_integer_to_chars(32) == chr(63 + 1) + chr(63)
assert encode_integer_to_chars(1023) == chr(63 + 31) + chr(63 + 31)
assert encode_integer_to_chars(1048575) == chr(63 + 31) * 6
assert encode_integer_to_chars(33) == chr(63 + 1) + chr(63 + 1)
```

These tests cover scenarios from zero to a number requiring more than one chunk, ensuring the function operates as expected.

## Example 2

### Instruction
Write a Python function to format a list of tag objects (each with a `.name` attribute) and into a single string containing all tag names separated by commas (no spaces), where tag names containing commas are first enclosed in double quotes.
The main function name is `format_tags`.

### Response
We will organize our response by four parts: [Reasoning], [Implementation], [Explanation], and [Tests].
\end{Verbatim}
In the prompt described above, we first craft system instructions to guide the model in generating high-quality code and tests while adhering to a specified format for easier parsing. From a pool of 10 high-quality examples \citep{wei2024selfcodealign}, we select one example to serve as the context for the generation. As discussed in Section \ref{sec:method}, we include the entry point in the instruction (i.e., \texttt{The main function name is format\_tags.}) to prevent compiler errors caused by mismatched entry points. Additionally, we conclude the prompt with the directive sentence \texttt{We will organize our response into four parts: [Reasoning], [Implementation], [Explanation], and [Tests].} This sentence ensures the model follows the required format, facilitating the parsing of code snippets and tests.

\end{document}